\documentclass[showpacs]{revtex4}
\usepackage{amssymb}
\usepackage{graphicx}
\usepackage{dcolumn}
\usepackage{bm}

\input{tcilatex}

\begin{document}

\title{The non-relativistic limit of (central-extended) Poincar\'{e} group
and some consequences for quantum actualization.}
\author{Juan S. Ardenghi and M. Castagnino}
\affiliation{Instituto de Astronom\'{\i}a y F\'{\i}sica del Espacio,\\
CC67 suc. 26 (1428) Buenos Aires, Rep. Argentina}
\email{ingatsac@iafe.uba.ar}
\author{R. Campoamor-Stursberg}
\affiliation{I.M.I.- Universidad Complutense de Madrid,\\
3 plaza de Ciencias, E-28040 Madrid}
\email{rutwig@pdi.ucm.es}

\begin{abstract}
The non relativistic limit of the centrally extended Poincar\'{e} group is
considered and their consequences in the Modal Hamiltonian Interpretation of
Quantum Mechanics discussed \cite{LC}, \cite{CL}. Through the assumption
that in Quantum Field Theory the Casimir Operators of the Poincar\'{e} Group
actualize, the non-relativistic limit of the latter group yields to the
actualization of the Casimir operators of the Galilei Group, which is in
agreement with the actualization rule of previous versions of modal
Hamiltonian Interpretation \cite{ACL}.
\end{abstract}

\maketitle

\section{Introduction}

This paper is mainly devoted to find the non-relativistic limit of an
extended Poincar\'{e} group. But as the motivation of this research was
inspired by a new interpretation of quantum mechanics \ we begin this
introduction discussing this subject.

For a long time after its first formulation, the interpretation of Quantum
Mechanics was deeply tied to possible measurement outcomes. Due to some
technical difficulties arising from this orthodox interpretation, new
approaches intended to give a more realistic description were worked out,
resulting in some interpretations that emphasize certain aspects or
properties of quantum systems, and establishing sharply the limits of the
corresponding approach.

\smallskip

In this sense, the modal interpretations first proposed by van Fraasen \cite%
{van1} establish a succinct distinction between the quantum state and the
value state of a system. While the former describes the possible physical
properties of the system, the latter represents the properties that can
actually be detected.

\medskip

Clearly this leads to a probabilistic relation among these concepts,
separated into two precise categories:

1.- The category of wave functions and density matrices, i. e., the world of
probabilities and potential facts.

2.- The category of actual facts, i. e., facts appearing in real
measurements. Examples are given when a dot appears in a photographic plate
or a Geiger counter detects the presence of a particle (facts that are
usually considered to be related with the quantum collapse).

In this context, the notion of ``actualization" serves as the link between
the two preceding classes. Actualization can be seen as the process in which
a potential fact from 1. becomes an actual fact. Obviously this procedure is
not free from some difficulties and constraints. The Kochen-Specker theorem (%
\cite{kochen},\cite{Bub}) specifies that not all the observables do
actualize, but only some of them. The various interpretations of Quantum
Mechanics are therefore committed to select a context in which the
observables that will acquire a definite value are defined. Thus, for
example, the Copenhagen interpretation establishes that only observables of
the measurement apparatus actualize, while for De Broglie-B\"ohm
interpretation the position observables actualize. The concept of
actualization has therefore become a crucial tool to interpret Quantum
Mechanics.

Enlarging the previous ans\"atze, in \cite{LC} a new member of the family of
Quantum Mechanics interpretations was introduced, the so-called Modal
Hamiltonian Interpretation (MHI). In this case, the Hamiltonian of the
quantum system plays a central role in the actualization problem. More
specifically, it is postulated that the observables corresponding to the
eigenbasis of a complete set of commuting observables containing the
Hamiltonian $H$ and the constants of motion must actualize. This approach is
therefore heavily based on the symmetry properties. In successive
refinements of the MHI \cite{CL}, it was proved that it agrees with the
``Ithaca interpretation" of Quantum Mechanics proposed by Mermin in \cite{MD}%
. We recall that Mermin's approach is based on various desiderata or
necessary conditions that any reasonable interpretation of Quantum Mechanics
should satisfy \cite{MD}.

Following the lines of this MHI, it was further proved in \cite{ACL} that
for systems invariant under the Galilei group, the three Casimir operators
of the central mass extended Galilei group, namely, the mass, the spin
squared, and the internal energy $W$ do actualize. Finally, the existence of
a kinetic component in the Hamiltonian $H$ was proved. Eliminating this
fictitious component $W\rightarrow H$, the postulates of \cite{LC} were
easily recovered.

In this paper we present a alternative version of the non-relativistic limit
of the centrally extended Poincar\'{e} group and its consequences in the
interpretation problem. We will also consider the non-relativistic limit of
the (centrally extended) Poincar\'{e} group as a necessary preparation to
implement the MHI for Quantum Field Theory (QFT). As this theory is
essentially a scattering theory, we will focus mainly on the preparation and
measurements stages. Nevertheless, we will consider the interaction stage,
which has no direct experimental verification. Therefore, at this stage, our
conclusions will be merely hypothetical. Several features of the problem are
also considered.

Section II presents some know features concerning the extended Galilei and
Poincar\'{e} Groups, developing in section III the non relativistic limit of
the trivially extended Poincar\'{e} group in both the classical and group
theoretical counterparts. In section IV we analyze the behavior of the
Casimir Operators under this limit. In section V be will consider the
particle number operator. Then in section V we will present our example:
following the ideas of \cite{ACL}) we show that, if we postulate that in QFT
the Casimir operators of the Poincar\'{e} group actualize, then the
non-relativistic limit of the latter group yields the actualization of the
Casimir operators of the Galilei group (namely those that actualize
according to the previous version of MHI of paper \cite{ACL}). In this
section we will essentially consider the one particle case. Finally, in
section VI, we consider the particle number that will also actualize, in the
preparation and measurement stages, since it labels different
representations of the corresponding groups. This section introduces the
many particles case. We will also discuss the difference between the in and
out stages and the interaction one and we will discuss some features of an
eventual version of MHI for QFT, introduce the preparation and measurement
apparatuses in detail, and we will consider also the corresponding
actualization. Section VII contains our conclusions.

\section{The (extended) Galilei and Poincar\'{e} groups}

In what follows, we will use the kinematical bases for both the Galilei and
Poincar\'e groups, as well as their extensions.

For the generators of the Galilei group we take the operators $%
H,P_{i},J_{i,}K_{Gi}$, where the last ones correspond to the Galilean
boosts. In this basis, the commutation relations are given by 
\begin{eqnarray}
\lbrack P_{i},P_{j}] &=&0,\;[J_{i},J_{j}]=\mathrm{i}\varepsilon
_{ijk}J_{k},\;[J_{i},P_{j}]=\mathrm{i}\varepsilon
_{ijk}P_{k},\;[H,P_{i}]=0,\;[H,J_{i}]=0,  \nonumber \\
\lbrack K_{Gi},K_{Gj}] &=&0,\;[J_{i},K_{Gj}]=\mathrm{i}\varepsilon
_{ijk}K_{Gk},\;[H,K_{Gi}]=-\mathrm{i}P_{i},\;[P_{i},K_{Gj}]=0,  \label{Gal1}
\end{eqnarray}

where $i,j,k,...=1,2,3$ and $\mathrm{i}=\sqrt{-1}$. As is well known, this
group admits a nontrivial central extension by a central charge $M$ that
commutes with the generators of the Galilei group. The brackets of the
extension are those of (\ref{Gal1}), with the exception of $[P_{i},K_{Gj}]=0$%
, that is replaced by

\begin{equation}
\lbrack P_{i},K_{Gj}]=-\mathrm{i}\delta _{ij}M  \label{r1}
\end{equation}%
While for an ordinary presentation (or at the classical level) this
extension is unnecessary, for quantum representations with an arbitrary
phase (i.e., such that $\left\vert \phi \right\rangle \sim \exp \left( {%
i\omega }\right) \left\vert \phi \right\rangle $\textbf{\ }) this central
extension \ is unavoidable (\cite{HW}, \cite{LB} chapter 3). Then the
extended group is the product of the primitive Galilei Group by a boolean
group with generator $M$.

\medskip

The generators of the Poincar\'{e} group are taken as: $H,P_{i},J_{i,}K_{Pi}$%
, where the last ones are the Lorentz boosts. The commutation relations of
the Poincar\'{e} group can be formulated in 4-dimension Lorentz space as:

\begin{eqnarray}
\lbrack P_{\mu },P_{\nu }]=0,\quad \lbrack M_{\mu \nu },P_{\rho }]=\mathrm{i}%
\left(\eta _{\mu \rho }P_{\nu }-\eta _{\nu \rho }P_{\mu }\right),  \nonumber
\\
\lbrack M_{\mu \nu },M_{\rho \sigma }]=\mathrm{i}\left(\eta _{\mu \rho
}M_{\nu \sigma }-\eta _{\mu \sigma }M_{\nu \rho }-\eta _{\nu \sigma }M_{\mu
\rho }+\eta _{\nu \rho }M_{\mu \sigma}\right),  \label{1.0}
\end{eqnarray}
where $\mu ,\nu ,...=0,1,2,3$ and 
\[
P_{\mu }=(H,P_{i}),\text{ }M_{\mu \nu }=\left( 
\begin{array}{cc}
0 & K_{Pi} \\ 
-K_{Pi} & J_{ij}%
\end{array}%
\right) , 
\]

$\eta _{\mu \nu }$ being the metric tensor of space-time, $J_{k}=\frac{1}{2}%
\varepsilon _{kij}J_{ij},$ and therefore $M_{\mu \nu }$ is defined by $%
(K_{Pi},J_{k})$. Then equations (\ref{1.0}) can be rewritten as follows: 
\begin{eqnarray}
\lbrack P_{i},P_{j}] &=&0,\;[J_{i},J_{j}]=\mathrm{i}\varepsilon
_{ijk}J_{k},\;[J_{i},P_{j}]=\mathrm{i}\varepsilon
_{ijk}P_{k},\;[H,P_{i}]=0,\;[H,J_{i}]=0,  \nonumber \\
\lbrack K_{Pi},K_{Pj}] &=&-\mathrm{i}\varepsilon _{ijk}J_{k},\
[J_{i},K_{Pj}]=\mathrm{i}\varepsilon _{ijk}K_{Pk},\ [H,K_{Pi}]=-\mathrm{i}%
P_{i},\ [P_{i},K_{Pj}]=-\mathrm{i}\delta _{ij}H.  \label{Poi}
\end{eqnarray}

\bigskip It follows at once from (\ref{Gal1}) and (\ref{Poi}) that the
Poincar\'{e} and Galilei groups share a splittable seven dimensional
subgroup $ISO(3)\times \left\langle H\right\rangle $ generated by $%
H,P_{i},J_{i,}$ and having commutators 
\begin{equation}
\lbrack P_{i},P_{j}]=0,\;[J_{i},J_{j}]=\mathrm{i}\varepsilon
_{ijk}J_{k},\;[J_{i},P_{j}]=\mathrm{i}\varepsilon
_{ijk}P_{k},\;[H,P_{i}]=0,\;[H,J_{i}]=0,  \label{1.1}
\end{equation}%
where $i,j,k,...=1,2,3$ \footnote{%
Generators $J_{i}$ define the $SO(3)$ group, generators $P_{i},J_{i,}$
define the $ISO(3)$ group: the inhomogeneous rotation group in three
dimensions. Actually (\ref{1.1}) is the largest subgroup that remains
invariant by the In\"{o}n\"{u}-Wigner contraction of the Poincar\'{e} on the
Galilei group.}.

\section{Extended Galilei group as a limit of the Poincar\'{e} group.}

It is well known that the Galilei group can be recovered from the Poincar%
\'{e} group by means of an In\"{o}n\"{u}-Wigner contraction \cite{LL}. It is
therefore natural to ask whether such a situation can be generalized to the
centrally extended Galilei group, which is one of the relevant objects in
Quantum Mechanics. However, since the Poincar\'{e} group does not admit
non-trivial central extensions \cite{Car}, the situation is more involved.
To solve this problem, we will consider two limiting processes: the first
using the classical reasoning of Special Relativity, and the second through
a special In\"{o}n\"{u}-Wigner contraction from a trivial extension of the
Poincar\'{e} group \footnote{%
By a trivial extension of a Lie algebra $\mathfrak{g}$ we mean the direct
sum $\mathfrak{g}\oplus M$, where $M$ is an additional commuting generator.}.

\subsection{The classical non-relativistic limit}

We perform the first approach using the non-relativistic limit in its usual
way. Let the position be given by $x^{\mu }=(ct,x^{i})$, the absolute
velocity be $U^{\mu }=(c,v^{i})$, the absolute time be $\tau =[t^{2}-\frac{1%
}{c^{2}}(x^{2}+y^{2}+z^{2})]^{\frac{1}{2}}$, the absolute mass be $m_{0}$
and the relative mass be $m=\gamma m_{0},$ where $\gamma =(1-\beta ^{2})^{-%
\frac{1}{2}}$ and $\beta =\frac{v}{c}.$ Then, if $p^{i}=mv^{i}$, the
absolute momentum reads $P^{\mu }=(\frac{E}{c},p^{i})$, where the energy is
given by $E=c^{2}m=c^{2}\gamma m_{0}$. We obtain the non-relativistic limit
for either $v\rightarrow 0$ or $c\rightarrow \infty $, namely, if $\beta
^{2}\rightarrow 0$. Expanding $\gamma $ into powers of $\beta $, we get 
\[
\gamma =(1-\beta ^{2})^{-\frac{1}{2}}=1+\frac{1}{2}\beta ^{2}+o(\beta ^{2}). 
\]%
If we neglect the quadratic term $o(\beta ^{2})$ below, \textit{i.e.}
physically we are considering velocities that are small compared to the
velocity of light, then 
\begin{equation}
P^{\mu }=(\frac{E}{c},p^{i})=(c\gamma m_{0},\gamma m_{0}v^{i})=(1+\frac{1}{2}%
\beta ^{2})(cm_{0},P_{G}^{i})=(cm_{0}+\frac{c}{2}\beta ^{2}m_{0},\text{ }%
P_{G}^{i}+\frac{1}{2}\beta ^{2}P_{G}^{i}),  \label{3.1}
\end{equation}%
where $P_{G}^{i}=m_{0}v^{i}$ is the non-relativistic momentum (the $P_{i}$
of the Galilei group). Now the term $cm_{0}$ can be promoted to an operator,
precisely the central charge $cM$. Taking into account the non-relativistic
limit $\beta ^{2}\rightarrow 0$, we deduce that 
\begin{equation}
\frac{E}{c}=\frac{H}{c}=cm=cm_{0}+\frac{c}{2}\beta ^{2}m_{0}=cM+\frac{T}{c}%
\rightarrow cM,  \label{3.2}
\end{equation}%
where $T=\frac{1}{2}m_{0}v^{2\text{ }}$ is the classical kinetic energy.
Taking the normalization $c=1$ (as in the commutation relations) in the
limit $\beta \rightarrow 0$, the latter expression reduces to 
\[
H\rightarrow M. 
\]%
Then (\ref{r1}) is the non relativistic limit of $\left[ {P_{i},K_{P_{j}}}%
\right] =-\mathrm{i}\delta _{ij}H$.

Let us finally observe that, from equation (\ref{3.2}), the proper mass $%
m_{0}$ can be considered, in the one particle case that we are considering
(see section 6 for the many particles case), as the internal energy $W$,
since it is the energy at the center of mass system.

Moreover, it follows from equation (\ref{3.1}) for $\beta \rightarrow 0$
that $p^{i}=P_{P}^{i}\rightarrow P_{G}^{i}$, i.e., the relativistic Poincar%
\'{e} momenta go to the non-relativistic Galilei momenta. Finally as the
Lorentz boost is given by 
\[
x^{\prime }=\gamma (x+vt),\;t^{\prime }=\gamma (t+\beta \frac{x}{c}) 
\]%
and, in the $\beta \rightarrow 0$ limit, this Galilei boost reads \footnote{%
See also in [11] for a $c=1$ deduction.} 
\[
x^{\prime }=x+vt,\;t^{\prime }=t 
\]%
Using these equations, it is easy to deduce that the relativistic boosts do
not commute, while the non-relativistic ones do. In this sense,$\left[ {%
K_{G_{i}},K_{G_{j}}}\right] =0$ is the non relativistic limit of the
commutator $\left[ {K_{G_{i}},K_{G_{j}}}\right] =-\mathrm{i}\varepsilon
_{ijk}J_{k}$ . This means that the Galilei group is the non relativistic
limit of the Poincar\'{e} group.

\subsection{Trivially extended Poincar\'{e} group and the generalized In\"{o}%
n\"{u}-Wigner contraction.}

In this paragraph we show that the previous non-relativistic limit can also
be justified by pure group theoretical arguments. For this purpose, we
reorder the generators of the Poincar\'{e} algebra $ISO(1,3)$ in the
following way: 
\[
\;[J_{i},J_{j}]=\mathrm{i}\varepsilon _{ijk}J_{k},\;[J_{i},K_{Pj}]=\mathrm{i}%
\varepsilon _{ijk}K_{Pk},[J_{i},P_{j}]=\mathrm{i}\varepsilon
_{ijk}P_{k},\;[H,J_{i}]=0, 
\]%
\begin{equation}
\lbrack P_{i},P_{j}]=0,[K_{Pi},K_{Pj}]=-\mathrm{i}\varepsilon _{ijk}J_{Pk},\
[H,K_{Pi}]=-\mathrm{i}P_{i},\ [P_{i},K_{Pj}]=-\mathrm{i}\delta
_{ij}H,\;[H,P_{i}]=0,  \label{4.1}
\end{equation}%
where in the l.h.s. of the first line we have listed all the commutator with 
$J_{i}$, which are related to the space isotropy. We now extend the group
trivially, i. e., in such a way that all the generators of (\ref{4.1})
commute with a trivial central charge $M$. Then we have a new algebra $%
I^{M}SO(1,3)=ISO(1,3)\times \langle M\rangle $ with basis $%
\{J_{i},P_{i},K_{Pi},H,M\}$. We perform the following change of the
generators' basis: 
\[
\overline{H}=H-M. 
\]%
Over the new basis $\{J_{i},P_{i},K_{Pi},\overline{H},M\}$, all commutators
of (\ref{4.1}) remain the same, with the only exception of 
\[
\lbrack P_{i},K_{Pj}]=-\mathrm{i}\delta _{ij}H=-\mathrm{i}\delta _{ij}(%
\overline{H}+M) 
\]

Observe in particular that space isotropy is preserved. We claim that this
algebra contracts naturally onto the centrally extended Galilei algebra
given by (\ref{Gal1}) where the last commutator is replaced by (\ref{r1}).
More specifically, the contraction 
\[
ISO(1,3)\times \langle M\rangle \rightsquigarrow G(2)\times \langle M\rangle 
\]%
is determined by the rescaling transformations (over the basis $%
\{J_{i},P_{i},K_{Pi},\overline{H},M\}$) defined by 
\begin{equation}
J_{i}^{\prime }=J_{i},\text{ }P_{i}^{\prime }=\varepsilon P_{i},\text{ }%
K_{Pi}^{\prime }=\varepsilon K_{Pi},\text{ }\overline{H}^{\prime }=\overline{%
H},\text{ }M^{\prime }=\varepsilon ^{2}M.  \label{4.4}
\end{equation}

It is straightforward to verify that the space isotropy remains unchanged by
this change of basis. The remaining commutation relations change as follows: 
\[
\lbrack P_{i}^{\prime },P_{j}^{\prime }]=0,\text{ }[K_{Pi}^{\prime
},K_{Pj}^{\prime }]=-\mathrm{i}\varepsilon ^{2}\varepsilon
_{ijk}J_{Pk}^{\prime },\ [H,K_{Pi}]=-\mathrm{i}P_{i}^{\prime },\ [H^{\prime
},P_{i}^{\prime }]=0, 
\]%
and 
\[
\lbrack P_{i}^{\prime },K_{Pj}^{\prime }]=-\mathrm{i}\delta _{ij}\varepsilon
^{2}(\overline{H}+M)=-\mathrm{i}\delta _{ij}(\varepsilon ^{2}\overline{H}%
^{\prime }+M^{\prime }). 
\]%
For the limit $\varepsilon \rightarrow 0$, the non-vanishing commutators are 
\[
\begin{array}{l}
\lbrack P_{i}^{\prime },P_{j}^{\prime }]=0,\text{ }[K_{Pi}^{\prime
},K_{Pj}^{\prime }]=0,\ [H^{\prime },P_{i}^{\prime }]=0, \\ 
\lbrack H^{\prime },K_{Pi}^{\prime }]=-\mathrm{i}P_{i}^{\prime },\
[P_{i}^{\prime },K_{Pj}^{\prime }]=-\mathrm{i}\delta _{ij}M^{\prime }.%
\end{array}%
\]%
It follows at once that the contracted algebra is isomorphic to the
extension of the Galilei algebra given by (\ref{Gal1}) where $\left[ {P_i
,K_{G_j } } \right] = 0 $ is replaced by (\ref{r1}).

As a consequence, the classical ansatz of the non-relativistic limit of the
Poincar\'{e} group inherits a group theoretical meaning, in terms of
generalized contractions of Lie algebras.

\section{Behavior of Casimir operators under non-relativistic limits}

We will now consider the non-relativistic limit of the Casimir operators.

The Casimir operators of the Poincar\'{e} group are well known \cite{LL},
and can be chosen as: 
\begin{equation}
\begin{array}{l}
C_{2}^{P}=H^{2}-P_{i}P_{i}, \\ 
C_{4}^{P}=H^{2}J_{i}J_{i}-(P_{i}P_{i})(K_{Pj}K_{Pj})+(J_{i}P_{i})^{2}-
(P_{i}K_{Pi})^{2}-2HJ_{k}\varepsilon ^{ijk}P_{i}K_{Pj}%
\end{array}
\label{13'}
\end{equation}%
over the kinematical basis. Their physical interpretation can be obtained in
the center of mass system, where $P_{i}=0$ and $H=E=m_{0}$ \footnote{%
For the same reasons explained in paper [3], where the transformation $%
H\rightarrow W$ is introduced, and the non-relativistic case analyzed.}. It
follows that 
\begin{equation}
C_{2}^{P}=m_{0}^{2},\;C_{4}^{P}=m_{0}^{2}J_{i}J_{i},  \label{5.1}
\end{equation}%
and we conclude that the two Casimir operators define the mass squared and
the spin squared, where the mass $m_{0}$ will be considered as the internal
energy in the one particle case. As observed earlier, we postulate that
these operators actualize. This postulate is based on experimental grounds,
since the mass and the spin of elementary particles are always well defined
in the preparation and measurement stages of a scattering process.

\medskip On the other hand, the Casimir operators of the trivially extended
Poincar\'{e} group $ISO(1,3)\times \langle M\rangle $, over the basis $%
\{J_{i},P_{i},K_{Pi},\overline{H},M\}$, are given by (see Appendix) 
\[
C_{1}^{PE}=M,\text{ \ \ }C_{2}^{PE}=-(P_{i}P_{i})+\overline{H}^{2}+M^{2}+2%
\overline{H}M, 
\]%
\begin{equation}
C_{4}^{PE}=(J_{i}J_{i})(\overline{H}%
+M)^{2}-(J_{i}P_{i})^{2}-(P_{i}P_{i})(K_{Pi}K_{Pi})+(P_{i}K_{Pi})^{2}-2(%
\overline{H}+M)\varepsilon _{ijk}J_{k}P_{i}K_{Pj}.  \label{5.11}
\end{equation}%
We observe that the change of basis in $ISO(1,3)\times \langle M\rangle $
explicitly introduces $\overline{H}$ and $M$ into the non-central Casimir
operator in a non-trivial way. In the centre of mass system ($P_{i}=0)$
these operators simplify to 
\begin{equation}
C_{1}^{PE}=m_{0},\;C_{2}^{PE}=m_{0}^{2},\;C_{4}^{PE}=m_{0}^{2}J_{i}J^{i}.
\label{15'}
\end{equation}%
As in the previous case, when $P_{i}=0$, the second Casimir operator
simplifies to $C_{2}^{PE}=(\bar{H}+M)^{2}=H^{2}$, showing that the
non-relativistic limit of the Hamiltonian is $H$ $\rightarrow $ $M$ (see
section 4.1). It is easily seen that the Casimir operators of the last
equation coincide with the operators of equation (\ref{5.1}), with the same
interpretation, to which $C_{1}^{PE}$ is added, the latter simply
corresponds to the mass.

\medskip

Proceeding in analogous manner, it is straightforward to see that the
Casimir operators of the mass central extended Galilei group are%
\begin{equation}
\begin{array}{l}
C_{1}^{G}=M=m_{0}, \\ 
C_{2}^{G}=ME-\frac{P^{2}}{2}=M(H-\frac{P^{2}}{2m_{0}})=m_{0}W, \\ 
C_{4}^{G}=M^{2}J_{i}J_{i}-(P_{i}P_{i})(K_{Gi}K_{Gi})+(P_{i}K_{i})^{2}-2MJ_{k}\varepsilon ^{ijk}P_{i}K_{Gj}, %
\label{5.2}%
\end{array}%
\end{equation}%
where $W$ is the internal energy. In the center of mass system these
operators have the form: 
\begin{equation}
C_{1}^{G}=m_{0},\;C_{2}^{G}=m_{0}^{2},\;C_{4}^{G}=m_{0}^{2}J_{i}J^{i}
\label{17'}
\end{equation}%
They therefore inherit the interpretation of proper mass, the proper mass
squared and the spin squared by the mass squared, coinciding with the
Casimir of equation (\ref{15'}).

\subsection{The limits among the Poincar\'{e} and Galilei groups.}

Following the previous section, the non-relativistic limit of the Casimir
operators of the Poincar\'{e} group equation (\ref{5.1}) onto the Casimir
operators of the extended Galilei group (\ref{17'}) can be obtained in the
following way:

\begin{enumerate}
\item $C_{2}^{P}=H^{2}-P_{i}P_{i}=m_{0}^{2}$ is the proper mass squared,
and, in the one particle case $W^{2}$, namely $C_{1}^{G2}$ or\ $C_{2}^{G}$
(cf. equation (\ref{17'})).

\item If we restore the $c$ factors in equation (\ref{13'}) we have that 
\[
C_{4}^{P}\sim
c^{-4}H^{2}J_{i}J_{i}-(P_{i}P_{i})(K_{Pi}K_{Pi})-(m_{0}J_{i}P_{i})^{2}+(P_{i}K_{Pi})^{2}-2c^{-2}HJ_{k}\varepsilon ^{ijk}P_{i}K_{Pj}. 
\]%
The term $(m_{0}J_{i}\beta _{i})^{2}$ becomes much smaller than the
remaining ones if $\beta _{i}\ll 1$. Therefore $C_{4}^{P}\rightarrow
C_{4}^{G}$ of equation (\ref{5.2}).
\end{enumerate}

We conclude that, in the center of mass system, the following limits hold: 
\[
C_{2}^{P}\rightarrow C_{2}^{G}=C_{1}^{G2},\;C_{4}^{P}\rightarrow C_{4}^{G}. 
\]%
So, in this case the two Casimir operators of the Poincar\'{e} group ($%
C_{2}^{P},$ $C_{4}^{P})$ go to the three Casimir operators of the extended
Galilei group ($C_{1}^{G},$ $C_{2}^{G},$ $C_{4}^{G})$. This anomaly, two
operators that go to three, is logically originated by the fact that \textit{%
we go from a non extended group to a extended one}. Nevertheless, we could
also say that somehow it is usual that two physical non-relativistic
entities become just one relativistic entity, e.g. space and time become
space-time. This fact would explain why the two Galilei Casimir operators $%
C_{1}^{G},$ $C_{2}^{G}$ (but really one since $C_{2}^{G}=(C_{1}^{G})^{2})$
become just one relativistic Poincar\'{e} Casimir $C_{2}^{P}.$

We now proceed to justify the preceding apparent anomaly by pure group
theoretical arguments, contracting the Casimir operators of the trivially
extended Poincar\'{e} Group onto those of the extended Galilei group. Using
the rescaled basis (\ref{4.4}) and expressing the Casimir operators, we
obtain that 
\[
\widetilde{C}_{1}^{PE}=\varepsilon ^{-2}M^{\prime },\text{ \ \ }\widetilde{C}%
_{2}^{PE}=-\varepsilon ^{-2}(P_{i}P_{i})+\overline{H}^{2}+\varepsilon
^{-4}M^{2}+2\varepsilon ^{-2}\overline{H}M, 
\]%
\[
\widetilde{C}_{4}^{PE}=(J_{i}J_{i})(\overline{H}+\varepsilon
^{-2}M)^{2}-\varepsilon ^{-2}(J_{i}P_{i})^{2}-\varepsilon
^{-4}(P_{i}P_{i})(K_{Pi}K_{Pi}) 
\]%
\[
+\varepsilon ^{-4}(P_{i}K_{Pi})^{2}-2\varepsilon ^{-2}\left( \overline{H}%
+\varepsilon^{-2}M\right)\varepsilon _{ijk}J_{k}P_{i}K_{Pj} 
\]%
The contracted Casimir operators are recovered, by the usual procedure, for $%
\varepsilon=0$: 
\[
\widehat{C}_{1}^{PE}=\lim_{\varepsilon \rightarrow 0}\varepsilon ^{2}%
\widetilde{C}_{1}^{PE}=M^{\prime },\text{ \ \ }\widehat{C}%
_{2}^{PE}=\lim_{\varepsilon \rightarrow 0}\varepsilon ^{4}\widetilde{C}%
_{2}^{PE}=M^{^{\prime }2} 
\]%
\begin{equation}
\widehat{C}_{4}^{PE}=\lim_{\varepsilon \rightarrow 0}\varepsilon ^{4}%
\widetilde{C}%
_{4}^{PE}=-(P_{i}P_{i})(K_{Pi}K_{Pi})+(P_{i}K_{Pi})^{2}+(J_{i}J_{i})M^{2}-2M%
\varepsilon _{ijk}J_{k}P_{i}K_{Pj}.  \label{5.12}
\end{equation}

From this equation and (\ref{5.2}), it follows at once that the non
relativistic limit is simply 
\[
\widehat{C}_{1}^{PE}=C_{1}^{G},\;\widehat{C}_{2}^{PE}=C_{2}^{G},\;\widehat{C}%
_{4}^{PE}=C_{4}^{G}. 
\]
Now we have three Casimir operators that go to three operators, because now
we go from an extended group to an extended group. \smallskip

So, there is a clearly better behavior if we proceed with an extended Poincar%
\'{e} Group for QFT, even if the extension is trivial. In fact, from
Wightman Axiom A we know that, in QFT, we deal with a Hilbert space composed
by normalized "rays", which are invariant under a phase transformation (see 
\cite{Haag}, page 58, or \cite{LB}, page 67). This kind of groups deserves
projective representations as in the extended groups above.

Then, from the non-relativistic limit of the Poincar\'{e} Casimir operators
that do \textit{actualize according to our postulate} in the introduction
(and well known physical facts), we deduce that, in the relativistic limit, 
\textit{the Galilei Casimir operators also actualize}, as proved in paper 
\cite{ACL} based on the postulates of MHI.

\section{The many particle case for the in and out stages.}

It is well known that in non-relativistic mechanics there are continuous
systems that cannot be conveniently modelled by a set of particles (solid,
waves, fluid, etc.). On the contrary, in QFT, at the in and out stages, all
systems can be modelled as a collection of $N$ particles (for simplicity we
will only consider sets consisting of a unique type of elementary particles)
and therefore the relevant group can be identified with the tensor product
of $N$-copies of the Poincar\'{e} group. It turns out that the
representations of this tensor product are expressible as products of
corresponding representations of the factor groups, so that these
representations are labelled by $N$ and the two (or three in the extended
case) Casimir operators of the Poincar\'{e} group. In this sense, the
particle number operator $N$ becomes an extra Casimir operator to be taken
into account.

\smallskip

Let the particle type be labelled by $C_{2}^{P}$ and $C_{4}^{P}$ or $%
C_{1}^{PE},$ $C_{2}^{PE}$ and $C_{4}^{PE}$ (as in section 5). Since a $n$
particle state is given by:

\[
|n\rangle =|1\rangle \otimes |1\rangle \otimes ...|1\rangle =(a^{\dagger
}\otimes a^{\dagger }\otimes ...a^{\dagger })|0\rangle =(a^{\dagger
})^{n}|0\rangle , 
\]%
where $a^{\dagger }$ is the creation operator of the particle we are
considering, and 
\[
N|n\rangle =n|n\rangle , 
\]%
the particle number is easily seen to be additive. Now we can combine the
already known Casimir operators (those of equation ( \ref{5.1})) and define
new operators for $N$-particles with the same mass and spin: 
\begin{equation}
\text{Mass}=(G_{2}^{P})^{\frac{1}{2}}N;\text{ \ \ \ Spin=}G_{4}^{P}\text{, \
\ Particle Number=}N.  \label{6.1}
\end{equation}%
Based on experimental facts, we state that these operators do actualize.

\smallskip

For the sake of completeness we could consider also the non-relativistic
limit of $N$, that would be the $N$ of non-relativistic quantum mechanics,
since the first is obtained from the $a^{\dagger }$ that creates the
relativistic particles, and the second obtained by its non-relativistic
limit, the $a^{\dagger }$ that corresponds to non-relativistic particles. Of
course, in the interaction stage we cannot consider the particle number
operator. We can only consider the Casimir operators of total mass, total
internal energy and total spin, that do actualize.

\section{Towards a Modal Hamiltonian Interpretation for QFT.}

In this section, in order to see how the previous results can be used, we
will anticipate some reasonings and consequences of a future possible
extension of MHI to QFT. We remember that the MHI belongs to the modal
family of interpretation, i. e. it is a realist, non collapse approach
according to which the quantum state describes the possible properties of
the system but not their actual properties (see \cite{LC} and \cite{CL}). In
this interpretation the Hamiltonian is essential for the definition of the
quantum systems and in the selection of its definite-valued observables that
may actualize, precisely, for MHI, they are observables that commute with
the Hamiltonian. In \cite{ACL} we show that the rule of definite-value
ascription, that selects a set of definite-valued observables must be
unaltered under the Galilean transformations. Then, since the Casimir
operators of the Galilean group are invariant under all the transformations
of the group, it is proposed that the actualization rule may be reformulated
in terms of these invariant operators. Moreover, in section 2 to 4, we have
shown that the Casimir operators have well behaved limits of the Casimir
operators of the extended Poincar\'{e} group. Then it is reasonable to
postulate that the Casimir operators of this extended Poincar\'{e} group
actualize in QFT, which is of course, universally admitted since this
Casimirs labels the representation of the group and therefore the type of
particles. But this is not the whole history since in QFT momenta actualize
in the in and out stages of the scattering theory. So in this section, in
order to verify these ideas about actualization in the simplest case, we
study the three stages of the scattering process of QFT, the only arena
where this theory has an experimental verification.

\subsection{In all three stages.}

Which observables actualize in QFT? As we have explained, $M$, $C_{2}^{PE}$
and $C_{4}^{PE}$ constitute the most natural choice for the three stages of
a QFT scattering experiment (and we know that they have a good
non-relativistic limit to the ordinary non-relativistic Quantum Mechanics).
The non-relativistic limit of this hypothetical choice yields the
non-relativistic MHI developed progressively in \cite{LC,CL,ACL}.

\subsection{In the in and out stages.}

Let us now consider the particular case of the in and out stages, where
experimentally we know that linear momenta are determined. Then we must also
explain the actualization of the components of the linear momentum, and it
could be possible that an additional postulate would be required. Actually
this is not the case, because in the preparation and measurement periods new
characters appear, namely the preparation and measurement apparatuses that
essentially define the momenta, in fact:

\begin{enumerate}
\item The preparation apparatus prepares a ray of particles with a well
defined momentum.

\item In the measurement apparatus as (e. g., in a fog or bubble chamber)
the momenta are well defined, (e. g. by trajectories in the chamber) since
their unique purpose is precisely to measure these momenta.

Then in principle in these three stages momenta are good candidates to
actualize. But let us try to deduce this conclusion directly from the
postulates of non-relativistic MHI, making the non-relativistic limit.
\end{enumerate}

\subsection{Non-relativistic limit}

In fact, to complete the panorama, we only need to consider the
non-relativistic limit of the preparation and measurement processes (via non
unitary evolutions) that begins and end in equilibrium states, related to
low velocities (which necessarily appears in these processes and correspond
to the non-relativistic case). This fact enables us to use the Galilei group
and its main features in the non-relativistic MHI.

Precisely:

\subsubsection{Preparation stage}

\begin{enumerate}
\item Theoretically, if we want to prepare an arbitrary state, we must begin
with a ground state and then accelerate it, using an external system, from
this initial state to the final state we want to reach (see for instance 
\cite{LB}, chapter 8).

\item The facts are the same in practice; particles initially at low
velocities must be accelerated.
\end{enumerate}

\subsubsection{Measurement stage}

Somehow you have the inverted process of the preparation stage.

\begin{enumerate}
\item Theoretically we must detect the trajectories of the outgoing
particles. Two main theoretical ways for this purpose are given:

\begin{enumerate}
\item We must introduce an environment, and then we have a composite system
formed by the elementary particle and the fog (or the bubbles) chamber.
Therefore we can explain the particle trajectory "\`{a} la Mott" \cite{NM}
(or using the Bohr-Oppenheimer method, already considered in \cite{LC} and 
\cite{CL}, since fog or the bubbles nuclei are larger than the scattered
particles and therefore are fixed).

\item According to the procedure developed in \cite{SID}, using the
destructive interference produced by the Hamiltonian evolution and showing
how classical trajectories appear. In all these cases, we necessarily reach
equilibrium, therefore previously we have low velocities that allow to use
the non-relativistic case of MHI.
\end{enumerate}

\item In practice, materializing these theoretical structures either with a
fog or bubble chamber or with a detector (Geiger counter, photographic
plate, etc.), we also have an irreversible and non unitary process to reach
equilibrium (general or partial but always with velocities small enough to
allow the recording of the results)
\end{enumerate}

So, either in the preparation or the measurement period, we have low
velocities and we can use (non-relativistic) MHI. But, the microscopical
objects, introduced by the preparation and measurement apparatuses,
necessarily produce some inhomogeneity and anisotropy \ in the total
Hamiltonian (scattered particles and apparatuses), and therefore they break
their eventual symmetries. We then arrive to the case of MHI (\cite{LC} and 
\cite{CL}), where the relevant Hamiltonian has few or no symmetries. Then,
there is a large number of constants of motion that introduce relevant
indices in the energy spectrum, and, according to MHI, all these constants
of motion actualize, among them the momenta $P_{i}$. Moreover, according to 
\cite{SID,SID2}, also well defined classical trajectories appear. Since
these trajectories are linear, they are defined by the corresponding momenta 
$P_{i}$, and this is a feature that proves that momenta do actualize.

In conclusion: in the relativistic case of QFT, based in our experience in
the non relativistic one and in obvious physical facts, we can postulate
that the Casimir operators do actualize, namely $C_{1}^{PE}$, $C_{2}^{PE}$
and $C_{4}^{PE}$, where, e. g., $C_{1}^{PE}$ is the mass of the proper
particle $m_{0}$ in the one particle case, and this mass multiplied by the
particle number $N$ is the total mass of the system in the many particles
case, in the in and out stages. Then the operators of equation (\ref{6.1})
are those that actualize. Finally, a possible explanation of the
actualization of linear momentum is given in this section.

This is just a sketch. We will follow this research hoping to find more
precise results.

\section{Conclusion}

We have presented the non relativistic limit of the trivially extended
Poincar\'{e} group: the extended Galilei group. We have shown that the
corresponding Casimir operators obey the same limit. Finally we have also
present an example where this limit can be used: the non relativistic limit
of a possible interpretation of QFT that turns out to be one of the possible
interpretations of Quantum Mechanics, MHI. This fact gives, at least, a
correct limit to our candidate interpretation for QFT. Moreover, in this
way, we can see how physical actualization changes according both to the
theory and the model considered and therefore may be we have found the base
for a modal interpretation of QFT.

\appendix

\section{Appendix}

The Casimir invariants can be computed using an analytical approach based on
the realization of Lie algebras in terms of differential operators \cite{BB}%
. For the extended Poincar\'e $ISO(1,3) \oplus \left\langle M \right\rangle $%
, the explicit system of PDEs obtained by this procedure is given by:

\begin{equation}
\begin{array}{l}
\widehat J_1 F = j_3 \frac{{\partial F}}{{\partial j_2 }} - j_2 \frac{{%
\partial F}}{{\partial j_3 }} + p_3 \frac{{\partial F}}{{\partial p_2 }} -
p_2 \frac{{\partial F}}{{\partial p_3 }} + k_3 \frac{{\partial F}}{{\partial
k_2 }} - p_2 \frac{{\partial F}}{{\partial k_3 }} = 0, \\ 
\widehat J_2 F = - j_3 \frac{{\partial F}}{{\partial j_1 }} + j_1 \frac{{%
\partial F}}{{\partial j_3 }} - p_3 \frac{{\partial F}}{{\partial p_1 }} +
p_1 \frac{{\partial F}}{{\partial p_3 }} - k_3 \frac{{\partial F}}{{\partial
k_1 }} - k_1 \frac{{\partial F}}{{\partial k_3 }} = 0, \\ 
\widehat J_3 F = j_2 \frac{{\partial F}}{{\partial j_1 }} - j_1 \frac{{%
\partial F}}{{\partial j_2 }} + p_2 \frac{{\partial F}}{{\partial p_1 }} -
p_1 \frac{{\partial F}}{{\partial p_2 }} + k_2 \frac{{\partial F}}{{\partial
k_1 }} - k_1 \frac{{\partial F}}{{\partial k_2 }} = 0, \\ 
\widehat P_1 F = p_3 \frac{{\partial F}}{{\partial j_2 }} - p_2 \frac{{%
\partial F}}{{\partial j_3 }} - \left( {\bar h + m} \right)\frac{{\partial F}%
}{{\partial k_1 }} = 0, \\ 
\widehat P_2 F = - p_3 \frac{{\partial F}}{{\partial j_1 }} + p_1 \frac{{%
\partial F}}{{\partial j_3 }} - \left( {\bar h + m} \right)\frac{{\partial F}%
}{{\partial k_2 }} = 0, \\ 
\widehat P_3 F = p_2 \frac{{\partial F}}{{\partial j_1 }} - p_1 \frac{{%
\partial F}}{{\partial j_2 }} - \left( {\bar h + m} \right)\frac{{\partial F}%
}{{\partial k_3 }} = 0, \\ 
\widehat K_1 F = k_3 \frac{{\partial F}}{{\partial j_2 }} - k_2 \frac{{%
\partial F}}{{\partial j_3 }} + \left( {\bar h + m} \right)\frac{{\partial F}%
}{{\partial p_1 }} - j_3 \frac{{\partial F}}{{\partial k_2 }} + j_2 \frac{{%
\partial F}}{{\partial k_3 }} + p_1 \frac{{\partial F}}{{\partial \bar h}} =
0, \\ 
\widehat K_2 F = - k_3 \frac{{\partial F}}{{\partial j_1 }} + k_1 \frac{{%
\partial F}}{{\partial j_3 }} + \left( {\bar h + m} \right)\frac{{\partial F}%
}{{\partial p_2 }} + j_3 \frac{{\partial F}}{{\partial k_1 }} - j_1 \frac{{%
\partial F}}{{\partial k_3 }} + p_2 \frac{{\partial F}}{{\partial \bar h}} =
0, \\ 
\widehat K_3 F = k_2 \frac{{\partial F}}{{\partial j_1 }} - k_1 \frac{{%
\partial F}}{{\partial j_2 }} + \left( {\bar h + m} \right)\frac{{\partial F}%
}{{\partial p_3 }} - j_2 \frac{{\partial F}}{{\partial k_1 }} + j_1 \frac{{%
\partial F}}{{\partial k_2 }} + p_3 \frac{{\partial F}}{{\partial \bar h}} =
0, \\ 
\widehat{\bar H}F = p_1 \frac{{\partial F}}{{\partial k_1 }} + p_2 \frac{{%
\partial F}}{{\partial k_2 }} + p_3 \frac{{\partial F}}{{\partial k_3 }} = 0,
\\ 
\widehat mF = 0. \label{a11}%
\end{array}%
\end{equation}

Here $\left\{ {j,p,j,\bar h,m} \right\}$ denote the coordinates of a basis
dual to the generators of the basis $\left\{J,P,K,\bar H,M\right\}$.

\begin{acknowledgments}
The three authors are extremely grateful to the referee, whose relevant
observations have greatly improved the final version of this paper. The
third author acknowledges partial financial support from the Research
Project No. MTM2006-09152 of the MICINN.
\end{acknowledgments}

\end{document}